\documentstyle[prb,aps,epsf]{revtex}

\newcommand{\Sp}{\mbox{{\bf s}}}

\newcommand{\be}{\begin{equation}}
\newcommand{\ee}{\end{equation}}
\newcommand{\ben}{\begin{eqnarray}}
\newcommand{\een}{\end{eqnarray}}
\newcommand{\ra}{\rangle}
\newcommand{\la}{\langle}
\newcommand{\im}{{\rm i}}
\newcommand{\Rg}{{\bf R}}
\newcommand{\kk}{{\bf k}}

\newcommand{\psbild}[1]{#1}  

\begin{document}

\twocolumn[\hsize\textwidth\columnwidth\hsize\csname@twocolumnfalse\endcsname

\title{Quantum Phase Transitions of a Square-Lattice\\
       Heisenberg Antiferromagnet with  Two Kinds of Nearest-Neighbour Bonds: \\
       A High-Order Coupled-Cluster Treatment}

\author{Sven E.~Kr\"uger, Johannes Richter, J\"org Schulenburg}
\address{Institut f\"ur Theoretische Physik,
         Universit\"at Magdeburg,\\
         P.O.Box 4120, D-39016 Magdeburg, Germany}
\author{Damian J.~J.~Farnell, Raymond F.~Bishop}
\address{Department of Physics, University of Manchester Institute of
         Science and Technology (UMIST),\\
         P.O.Box 88, Manchester M60 1QD, United Kingdom.}

\date{\today}
\maketitle

\begin{abstract}
We study the zero-temperature phase diagram
and the low-lying excitations of
a square-lattice spin-half Heisenberg antiferromagnet with two types of
regularly distributed
nearest-neighbour exchange bonds ($J>0$ (antiferromagnetic)
and $-\infty < J' <
\infty $) using the coupled cluster method (CCM) for high orders
of approximation (up to LSUB8). We use a N\'eel model state as well as a helical model 
state as a starting point for the CCM calculations.
We find a second-order transition from a phase with N\'eel order to 
a finite-gap quantum disordered phase for sufficiently large 
antiferromagnetic exchange constants $J'>0$. For 
frustrating ferromagnetic couplings $J'<0$
we find indications that quantum fluctuations favour 
a first-order phase transition from the
N\'eel order to a quantum helical state, by contrast with the 
corresponding second-order 
transition in the corresponding classical model. 
The results are compared to those of exact diagonalizations of finite systems
(up to 32 sites) and those of spin-wave and variational calculations. The CCM results
agree well with the exact diagonalization data
over the whole range of the parameters.
The special case of $J'=0$, which is equivalent to the honeycomb lattice, is treated
more closely.
\end{abstract}

]  

\section{Introduction}
 The subject of quantum spin-half
 antiferromagnetism in low-dimensional systems
 has attracted a great deal of interest in recent times in
 connection with
 the magnetic properties of the cuprate high-temperature
 superconductors. However, low-dimensional quantum spin systems are
 of interest in their own right  as examples of strongly interacting
 quantum  many-body systems.
 Although we know from the Mermin-Wagner theorem \cite{mermin66} that 
 thermal fluctuations are strong enough to
 destroy magnetic long-range order at any finite temperature, 
 the role of quantum fluctuations is less understood.
 As a result of intensive work in the late eighties, it is now
 well-established that the ground-state of the
 Heisenberg antiferromagnet on the square lattice
 with nearest-neighbour interactions is
 long-range ordered (see for example the review in Ref.~\onlinecite{manou91}).
 However, Anderson's and Fazekas'
 investigations \cite{anderson73} of the
 triangular lattice led to a conjecture that quantum fluctuations
 plus frustration may be sufficient to destroy the N\'eel-like
 long-range order in
 two dimensions.
 Another specific area of recent research is the spin-half
$J_1$--$J_2$ antiferromagnet on the 
square-lattice 
 where the frustrating diagonal $J_2$ bonds plus quantum fluctuations are 
able to realize a second-order transition from N\'eel ordering to a disordered
quantum spin liquid
(see for example Refs.~\onlinecite{ri93,chub94a,oitmaa96,Bishop6} and references
 therein). On the other hand, there are cases in which frustration causes a
first-order transition in quantum spin systems in contrast to a second-order
transition in the corresponding classical model (see for example 
Refs.~\onlinecite{xian95,niggemann97,richter98,ivanov98}).

 Besides frustration,
 there is another mechanism to realize the ``melting'' of N\'eel ordering in the
 ground states of unfrustrated Heisenberg antiferromagnets,
 namely the
 formation of local singlet pairs of two coupled spins.
 This mechanism may be relevant
 for the quantum disordered state in bilayer systems 
 \cite{millis_monien,sandvik_scalapino,gros_wenzel_richter,chubukov_morr} as 
well as in CaV$_4$O$_9$ (see for example Refs.~\onlinecite{troyer96,sachdev96,albrecht96}
and references therein).
 The formation of local singlets is connected with a gap in the
 excitation spectrum. By contrast, the opening of a gap in the excitation
 spectrum of frustrated
 systems seems to be less clear and might be dependent on details of the
 exchange interactions.

In the present paper, we study a model which contains both mechanisms,
frustration and singlet formation, in different parameter regions.
We mainly use in this article the coupled cluster method,\cite{coester58,bishop87,bishop91}
which has become widely recognized as 
one of the most powerful and most universal techniques in quantum many-body theory.
In recent years there has been increasing success in applying the CCM to
quantum spin 
systems,\cite{Bishop6,roger90,bishop91a,Bishop2,Bursill,Farnell2,rosenfeld98,zeng98,bishop99}
especially with the advent of high-order approximations
which utilize computer algebra.\cite{zeng98} Subsequently, high-order
CCM approximations have been applied to the $XXZ$ model,\cite{zeng98} the
anisotropic $XY$ model\cite{Farnell2} and the $J_1$--$J_2$ model.\cite{Bishop6}
In addition to the CCM results we also present variational,
spin-wave theory (SWT) and exact diagonalization (ED) results for the sake of comparison.

\section{The model}
\begin{figure}[h]
  \psbild{\centerline{\epsfysize=7cm \epsfbox{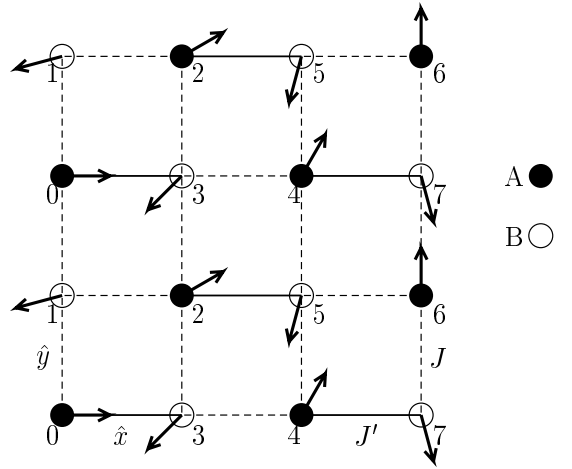}}}
  \caption{Illustration of the classical helical state for the square-lattice
           Heisenberg antiferromagnet of Eq.~(\protect{\ref{ham}}),
           with two kinds of regularly distributed nearest-neighbour
           exchange bonds, $J$ (dashed lines) and $J'$ (solid lines).
           The spin orientations at $A$ and $B$ lattice sites are defined
           by the angles $\theta_n=n\Phi$ and $\theta_n=n\Phi+\pi$,
           respectively, where $n=0, 1, 2, ...$, and $\Phi$ is the
           characteristic angle of the helical state.
           The state is shown for
           $\Phi=\pi/12$ and $n=0, 1,\dots, 7$.}
  \label{fig1}
\end{figure}

We consider a spin-half Heisenberg model on a square
lattice with nearest neighbor bonds $J$ and $J'$
in a regular zigzag pattern as shown in Fig.~\ref{fig1}. The
Hamiltonian is given by
\ben\label{ham}
        H & = & J\sum_{<ij>_1}\Sp_i\cdot\Sp_j+J'\sum_{<ij>_2}\Sp_i\cdot\Sp_j \nonumber \\
          & \stackrel{J=1}{=} & \sum_{i\in A}\sum_p(1+\delta_{p,p_{J'}}(J'-1))\Sp_i\cdot\Sp_{i+p} .
\een
The sums over $<ij>_1$, and $<ij>_2$ represent sums over the nearest-neighbour 
bonds shown in Fig.~\ref{fig1} by dashed and solid lines, respectively.
Throughout the paper we fix the $J$ bond to be antiferromagnetic ($J>0$)
and henceforth scale it to the value $J=1$,
and consider $J'$ as the free parameter of the model. 
We also split the square lattice 
into the equivalent $A$ and $B$ sublattices shown in Fig.~\ref{fig1}.
In Eq.~(\ref{ham})
the sum over $i$ runs over the sites of the sublattice $A$,
with vectors $p=\{(0,\pm 1),(\pm 1,0)\}$ connecting nearest neighbours. 
In particular, $p_{J'}=(1,0)$ pertains to the coupling with $J'$ bonds.

Each square lattice plaquette
consists of three $J=1$ bonds and one $J'$ bond. In the case of
ferromagnetic $J'$ bonds (i.e., $J'<0$), 
the plaquettes are frustrated. Conversely, for
antiferromagnetic $J'$ bonds (i.e., $J'>0$) 
there is no frustration in the system,
although the difference of the coupling strengths $J$ and $J'$
leads to quantum competition.
This model has been treated previously using perturbation theory,\cite{singh88}
renormalized spin wave theory (RSWT)\cite{ivanov96} and exact diagonalization
(ED).\cite{richter97}
It allows us to study the influence of local singlet
formation ($J' > 1$) and  frustration ($J' < 0$) on the stability of the 
N\'eel order within a single model.

Ferromagnetic bonds in an antiferromagnetic matrix have been discussed in
recent times \cite{ri93,schlottmann,betts93,rodriguez,korenblit} 
in connection with the proposal by
Aharony and coworkers \cite{aharony88} to model localized oxygen holes in the
Cu-O-planes by local ferromagnetic bonds
between the copper spins.
It was argued that random ferromagnetic bonds may influence the
antiferromagnetic order drastically and may support the realization of a
quantum spin-liquid state.\cite{ri93,rodriguez,korenblit}

On the other hand, the case of antiferromagnetic $J'$ bonds with
$J'>1$ resembles the situation in bilayer systems and in the depleted
square-lattice antiferromagnet CaV$_4$O$_9$, in which the
competition between two different antiferromagnetic bonds leads to a phase
transition from antiferromagnetic long-range order to quantum disorder with a
finite gap. It is seen in this article that the transition point obtained for the
model of Eq.~(\ref{ham}) is quite close 
to that obtained for the bilayer model.\cite{sandvik_scalapino}

There are some special cases of the model Hamiltonian of Eq.~(\ref{ham}):\\
(i) \hspace{0.2cm} $J'=1$: square-lattice antiferromagnet, for which the
ground state is long-range ordered; \\
(ii) \hspace{0.1cm}$J'=0$: honeycomb-lattice antiferromagnet, for which 
the ground state is  long-range ordered; \\
(iii) $J'=-\infty$: spin-$1$ triangular lattice, for which the ground
state is long-range ordered; and \\
(iv)\hspace{0.1cm} $J'=+\infty$: valence-bond solid, for which the ground
state is a rotationally invariant quantum dimer state with an excitation gap.

{\em Classical ground state.} 
For $J' >-1/3$ the N\'{e}el state is the classical ground state
of the Hamiltonian of Eq.~(\ref{ham}).
At  $J'_c=-1/3$
there is classically a second-order phase transition to a ground state of helical nature
(see Fig.~\ref{fig1}),
with a characteristic pitch angle $\Phi=\pm|\Phi_{\rm cl}|$ given by
\be\label{phi}
   |\Phi_{\rm cl}|=\left\{\begin{array}{ll}
                                                          0 & \quad J'>-\frac{1}{3} \\
    \arccos\left(\frac{1}{2}\sqrt{1-\frac{1}{J'}}\; \right) & \quad J'\le -\frac{1}{3} \\
    \end{array}\right.
\ee
where the different
signs correspond to the two chiralities \cite{villain} of this
helical state. Note that for $\Phi=0$ this is just the N\'eel state.
More generally, the 
pitch angle varies with $J'$ 
from $|\Phi_{\rm cl}|=0$ for $J'>-1/3$ to  $|\Phi_{\rm cl}|= \pi/3$
for $J'=-\infty$. Note
that $|\Phi_{\rm cl}|=\pi/3$ (realized at  $J'= -\infty$) corresponds to
the ground state of the spin-1 triangular lattice.
We describe the directions of the spins $\Sp_A$ and $\Sp_B$, belonging
to the $A$ and $B$ sublattices respectively, for the classical helical state
with a characteristic angle $\Phi$ as follows\cite{ivanov96} 
(and see Fig.~(\ref{fig1})),
\begin{eqnarray}\label{klass_spins}
     \Sp_A({\bf R}) &=&{\bf \hat{u}}\cos{\bf Q\cdot R}+{\bf \hat{v}}\sin{\bf Q\cdot R},
     \\
     \Sp_B({\bf R}+\hat{x}) &=&{\bf \hat{u}}\cos({\bf Q\cdot R}+\pi+3\Phi)
     +{\bf \hat{v}}\sin({\bf Q\cdot R}+\pi+3 \Phi), \nonumber
\end{eqnarray}
where ${\bf \hat{u}}$ and ${\bf \hat{v}}$ are perpendicular
unit vectors in the spin space, ${\bf R}$ runs over the sites of
the sublattice $A$, and we have ${\bf Q}=(2\Phi,0)$ for the
pitch vector ${\bf Q}$.
We note that this general
helical state does not have a periodicity in the $x$-direction because
$\Phi$ is in general not of the form $m\pi/n$ with $m$ and $n$ integral.
We also note that we have only three
different angles between nearest-neighbour spins, namely
$\pm(\pi-\Phi)$ for the $J=1$ couplings and $\pi-3\Phi$ for the coupling with
$J'$.

The maximum frustration is in the region around  $J'\approx -1$. 
Bearing in mind the situation for the $J_1$--$J_2$ model, one might expect that 
for the extreme quantum case (spin-half)
quantum fluctuations might be able to open the window
to a spin-liquid phase for a finite range of parameters around
this region of maximum frustration. On the other hand, for strong
antiferromagnetic bonds ($J' \gg 1$) there is, of course, 
no indication in the
classical model for the breakdown of the N\'eel order.

{\em Simple variational ansatz for the quantum ground-state.} 
In the quantum case, the  region of strong antiferromagnetic 
$J'$ bonds ($J' \gg 1$) is characterized by
a tendency to singlet pairing of the two spins corresponding to a $J'$ bond.
Using a high-order series expansion \cite{singh88} the 
N\'eel order was found to be stable up to a critical value $J'_s\approx 2.56$.
 
A comparable value can be obtained using
a simple variational wave
function similar to that used \cite{gros_wenzel_richter} for bilayer
systems, namely
\be \label{var} |\Psi_{\rm var}\rangle = \prod_{i \in A}
   \frac{1}{\sqrt{1+t^2}} \big[|\uparrow_i \downarrow_{i+\hat{x}}\rangle
   -t |\downarrow_i \uparrow_{i+\hat{x}}\rangle\big],
\ee
where the lattice sites $i$ and
$i+\hat{x}$ correspond to a $J'$ bond, and where the product in Eq.~(\ref{var})
is thus effectively taken over the $J'$ bonds of the lattice of Eq.~(\ref{ham}).
The trial function depends on the variational parameter $t$ and
interpolates between a valence-bond state realized for $t=1$ and the
N\'eel state for $t=0$. For $t=1$, the singlet pairing is complete and
$|\Psi_{\rm var}\rangle$ represents an eigenstate of the model
of Eq.~(\ref{ham}) in the limit $J'\rightarrow\infty$ (dimer state).
By minimizing $\la\Psi_{\rm var}|H|\Psi_{\rm var}\ra$ 
with respect to the variational parameter $t$ we get an upper bound
for the ground-state energy per spin of the model of Eq.~(\ref{ham}),
\be\label{en_var}
   E_{\rm var}/N =\left\{\begin{array}{ll}
         -(J'^2+3J'+9)/24  & \quad J'\le 3 \\
         -3J'/8            & \quad J'>   3 \\
    \end{array}\right.
.\ee
The relevant order parameter describing the N\'eel order is
\be\label{m_var}
   M_{\rm var}=\la\Psi_{\rm var}|s_i^z|\Psi_{\rm var}\ra = \left\{\begin{array}{ll}
         1/2\sqrt{1-J'^2/9} & \quad J'\le 3 \\
         0                  & \quad J'>   3 \\
    \end{array}\right.
,\ee
showing a breakdown of the N\'eel order at a critical value $J'_s=3$.

\section{Coupled cluster calculations}
\subsection{The ground-state formalism}
The starting point for any CCM calculation 
(see overwiev in Ref.~\onlinecite{bishop91})
is the choice of a normalized model or reference state $|\Phi\rangle$,
together with a set of mutually commuting
multispin creation operators $C_I^+$ which are defined over a complete set of
many-body configurations $I$. The operators
$C_I$ are the multispin destruction operators
and are defined to be the Hermitian
adjoints of the $C_I^+$. We choose $\{|\Phi\ra;C_I^+\}$ in such a way
that we have $\la\Phi|C_I^+=0=C_I|\Phi\ra$, $\forall I\neq 0$, where,
by definiton, $C_0^+=\openone$, the identity operator.

For spin systems, an appropriate choice for the CCM model state $|\Phi\ra$
is often a classical spin state,\cite{bishop91a} in which the most general
situation is that each spin can point in an arbitrary direction. 
For the case of the Hamiltonian of Eq.~(\ref{ham}), we choose 
the helical state illustrated in Fig.~\ref{fig1} to be our model state.
Although the classical ground state of Eq.~(\ref{ham})
is precisely of this form,
we do not choose the classical result for the pitch angle $\Phi$ 
but we consider it rather as a free parameter in the CCM calculation.

In order to perform a CCM calculation, we would like to treat each
site equivalently and we do this by performing a rotation of the local spin
axes at each site about the $y$-axis such that all spins in the model state
align in the same direction, say down (along the negative $z$-axis).
After this transformation we have 
\be 
   |\Phi\ra=|\cdots\downarrow\downarrow\downarrow\cdots\ra; \quad C_I^+=s_r^+,\,\, s_r^+s_{r'}^+,\,\, s_r^+s_{r'}^+s_{r''}^+,\cdots
,\ee
(where the indices $r,r',r'',\dots$ denote any lattice site) respectively,
for the model state and the multispin creation operators, which
now consist of spin-raising operators only.

In order to make the spin $\Sp_i$ point down let us suppose we need to perform
such a rotation of the spin axes by an angle $\delta_i$.
This is equivalent to the transformation
\ben\label{ks_trafo}
   s_i^x & \rightarrow & \cos\delta_i s_i^x+\sin\delta_i s_i^z \nonumber \\
   s_i^y & \rightarrow & s_i^y \\
   s_i^z & \rightarrow & -\sin\delta_i s_i^x+\cos\delta_i s_i^z \nonumber
.\een
A similar rotation about the $y$-axis by an angle
$\delta_j$ is performed for the spin $\Sp_j$. Thus we get for the
transformation of the scalar product of the two spins,
$\Sp_i\cdot\Sp_j \rightarrow (\Sp_i\cdot\Sp_j)_{\varphi}$, where
\ben\label{produkt_trafo}
   (\Sp_i\cdot\Sp_j)_{\varphi} &  \equiv & \sin\varphi[s_i^xs_j^z-s_i^zs_j^x]+\cos\varphi[s_i^xs_j^x+s_i^zs_j^z] + s_i^ys_j^y \nonumber \\
   & = &\frac{1}{2}\sin\varphi[s_i^+s_j^z-s_i^zs_j^++s_i^-s_j^z-s_i^zs_j^-] +\cos\varphi s_i^zs_j^z  \nonumber \\
   & &  +\frac{1}{4}(\cos\varphi+1)[s_i^+s_j^-+s_i^-s_j^+] \\
   & &  +\frac{1}{4}(\cos\varphi-1)[s_i^+s_j^++s_i^-s_j^-] \nonumber
.\een
The angle $\varphi\equiv\delta_j-\delta_i$ is the angle between the two spins, and
$s^{\pm}\equiv s^x\pm \im s^y$ are the spin-raising and spin-lowering operators.
Note that this product of two spins after the rotation
depends not only on the angle {\em between} them, but also on the
sign of this angle.
In case of the N\'eel model state ($\Phi=0$), the angle between any
neighbouring spins is $\pi$, and hence Eq.~(\ref{produkt_trafo}) becomes
$\Sp_i\cdot\Sp_j \rightarrow -\frac{1}{2}[s_i^+s_j^++s_i^-s_j^-]-s_i^zs_j^z$.

Using the helical state of Eq.~(\ref{klass_spins}) with the
characteristic angle $\Phi$, the Hamiltonian of Eq.~(\ref{ham}) is now
rewritten in the local coordinates as,
\be\label{ham2}
    H=\sum_{i\in A}\sum_p(1+\delta_{p,p_{J'}}(J'-1))(\Sp_i\cdot\Sp_{i+p})_{\varphi_p}  ,\ee
where the angles between neighbouring spins are
$\varphi_{\pm\hat y}=\pi+\Phi$, $\varphi_{-\hat x}=\pi-\Phi$ and $\varphi_{\hat x}=\pi+3\Phi$.
While the general helical state (see Fig.~\ref{fig1}) does not have
translational symmetry in the $x$-direction, the transformed Hamiltonian of Eq.~(\ref{ham2})
does have this symmetry since it depends only on the angles between neighbouring
spins.

Having defined an appropriate model state $|\Phi\ra$ with creation
operators $C_I^+$, the CCM parameterizations of the ket and bra ground 
states are given by
\be\label{ket} 
  |\Psi\ra =e^S|\Phi\ra, \quad S=\sum_{I\neq 0}{\cal S}_IC_I^+ ,\ee
\be\label{bra} 
  \la\tilde\Psi|=\la\Phi|\tilde Se^{-S}, \quad \tilde S=1+\sum_{I\neq 0}\tilde {\cal S}_IC_I .\ee
The correlation operator $S$ is expressed in terms of 
the creation operators $C_I^+$ and the ket-state correlation coefficients
${\cal S}_I$. 
We can now write the ground-state energy as,
\be\label{en} 
    E=\la\Phi|e^{-S}He^S|\Phi\ra .\ee
To describe the magnetic order of the system, we use a simple order parameter
which is expressed in terms of the local, rotated spin axes, 
and which is given by
\be \label{m}
 M\equiv-\la\tilde\Psi|s_i^z|\Psi\ra ,\ee
such that the order parameter represents the on-site magnetization.
Note that $M$ is the usual sublattice magnetization for the case of 
the N\'eel state as the CCM model state.

To find the ket-state and bra-state correlation coefficients we have
to require that the expectation value $\bar H=\la\tilde\Psi|H|\Psi\ra$
is a minimum with respect to the 
bra-state and ket-state correlation coefficients.
This formalism is exact if we include all possible multispin configurations in
the correlation operators $S$ and $\tilde S$, which 
is usually impossible in a practical situation.
We use the LSUB$n$ approximation scheme\cite{zeng98} to truncate the expansion
of $S$ and $\tilde S$ in the Eqs.~(\ref{ket}) and (\ref{bra}).

Using the lattice symmetries, we have now to find all
{\em different} possible configurations with respect to the point and space group 
symmetries of both the lattice and Hamiltonian
with up to $n$ spins
spanning a range of no more than $n$ adjacent
lattice sites (LSUB$n$ approximation) and 
these are referred to as the fundamental configurations.

The Hamiltonian of Eq.~(\ref{ham}) has four lattice point-group symmetries namely
two rotational operations ($0^{\circ}$, $180^{\circ}$) and two
reflections (along the $x$- and $y$-axes), defined by:
\ben x\rightarrow x, \quad y\rightarrow y;  & &\qquad  x\rightarrow -(x+1), \quad y\rightarrow -y, \nonumber \\
     x\rightarrow x, \quad y\rightarrow -y; & &\qquad  x\rightarrow -(x+1), \quad y\rightarrow y .
\een
The rotation of $180^{\circ}$ and the reflection along the $y$-axis are
connected by a shift of $\hat x=(1,0)$.
The translational operator $T$ is defined by
\be
   T=(n+m)\hat x+(m-n)\hat y, \quad \mbox{$n$, $m$ integral} ,\ee
such that translational symmetry is preserved.

The N\'eel model state also contains these symmetries, and so for
this model state we can directly apply all these symmetries in finding
the fundamental configurations.
On the other hand the general helical model state ($\Phi\neq 0$) has only two
of the above four lattice point-group symmetries,
namely $x\rightarrow x, \quad y\rightarrow y$ and
$x\rightarrow x, \quad y\rightarrow -y$, and so this reduced symmetry yields
a larger number of fundamental configurations.

In the case of the N\'eel model state ($\Phi=0$), the number of fundamental configurations
can further be reduced by explicitly conserving the total uniform magnetization
$s_T^z\equiv\sum_k s_k^z$ (the sum on $k$ runs over all lattice sites) because
the ground state is known to lie in the $s_T^z=0$ subspace.
This means that we exclude configurations with an odd number of spins, 
and therefore we do not use LSUB3, LSUB5, etc.~approximations.
The helical state is not
an eigenstate of $s_T^z$ and we cannot apply this property when using
the helical model state.
The fundamental configurations
can now be calculated computationally,\cite{zeng98} and
the resulting numbers of LSUB$n$ configurations for $n\leq 8$
are given in Table~\ref{tab1}.

The Ket-state and bra-state equations are calculated
computationally.\cite{zeng98}
For the N\'eel model state, we are able to carry out the CCM
up to the LSUB8 level (where we need to solve 4986 coupled equations),
whereas for the helical state we could do this only up to the LSUB6 level
(where we need to solve 1638 coupled equations).

\subsection{The excited state formalism}
We use the excited-state formalism of Emrich\cite{emrich81,bishop91a,bishop99a}
to approximate the excited-state wavefunctions. We apply
an excitation operator $X^e$ linearly to the ket state wavefunction (\ref{ket}),
such that 
\be \label{ket_ex} |\Psi_e\ra=X^ee^S|\Phi\ra; \quad X^e=\sum_{I\neq 0}{\cal X}_I^eC_I^+ .\ee
Using the Schr\"odinger equation, $H|\Psi_e\ra=E_e|\Psi_e\ra$, we
find that
\be \label{ket_exc} \epsilon_eX^e|\Phi\ra=e^{-S}[H,X^e]_-e^S|\Phi\ra ,\ee
where $\epsilon_e$ ($\equiv E_e-E$) is the difference between the
excited-state energy ($E_e$) and the ground-state energy ($E$).
Applying $\la\Phi|C_I$ to Eq.~(\ref{ket_exc}) we
find that,
\be \label{ew} \epsilon_e{\cal X}_I^e=\la\Phi|C_Ie^{-S}[H,X^e]_-e^S|\Phi\ra ,\ee
which is an eigenvalue equation with eigenvalues $\epsilon_e$ and
corresponding eigenvectors ${\cal X}_I^e$.

As for the ground state, we must use an approximation scheme for $X^e$ in
Eq.~(\ref{ket_ex}). Although it is not necessary\cite{bishop99a} to use
the same approximation for the excited state as for the ground state, we
in fact do so to keep the CCM calculations as systematic and self-consistent
as possible.
We define the fundamental configurations for LSUB$n$ (for the N\'eel state)
as previously, but we now restrict the choice of configurations to contain
only those which produce a change of $s_T^z$ of $\pm 1$ with respect
to the model state.\cite{bishop99a} Since we are only interested in the
lowest-lying excitations, the restriction to these single-magnon
spin-wave-like excitations is the correct choice.
The number of fundamental excited-state configurations
for LSUB$n$ is given in Table~\ref{tab1}.

To calculate the terms of the right hand side of Eq.~(\ref{ew})
we use the same computational
algorithm as for the calculation of the ground-state ket equation.
The terms contain the ground ket-state correlation coefficients ${\cal S}_I$,
so once these coefficients have been determined the eigenvalue equation
(\ref{ew}) can be solved (numerically).
We furthermore choose the lowest energy eigenvalue of Eq.~(\ref{ew}) in order
to calculate the excitation energy gap, $\Delta$. We note that the eigenvalues
of Eq.~(\ref{ew}) are not guaranteed to be real, since as a generalized
eigenvalue equation it is not symmetric. However, over the entire regime
of interest, the values of $\Delta$ so obtained are found to be real.
We have performed these calculations for the excited state up to the
LSUB6 level of approximation.

\subsection{Extrapolation of the CCM-LSUB$n$ results}
Although no scaling theory for results of LSUB$n$ approximations has yet
been proven, there are empirical indications\cite{Bishop2,Farnell2,zeng98,bishop99a}
of scaling laws for the energy, the magnetization, and the
excited-state energy gap for various spin models.

These scaling laws can be justified by the observations that they fit the
results well (i.e., with low mean-square deviation), and that the
extrapolated results are in good agreement with results of other methods
(e.g., Green function Monte Carlo or series expansion for the 2D
$XXZ$ model\cite{Bishop2,zeng98}) or
with exact results (e.g., 1D $XY$ model\cite{Farnell2}).
In accordance with those previous results we use the following
scaling laws: for the ground-state energy,
\be E=a_0+a_1(1/n^2)+a_2(1/n^2)^2 ;\ee
for the ground-state magnetization,
\be M=b_0+b_1(1/n)+b_2(1/n)^2 ;\ee
and for the gap of the lowest-lying excitations,
\be \Delta=c_0+c_1(1/n)+c_2(1/n)^2 ;\ee
where $n$ is the LSUB$n$ approximation level.

\subsection{Choice of the CCM model state}
As stated previously, we use the helical state of Eq.~(\ref{klass_spins})
with the characteristic angle $\Phi$, illustrated in Fig.~\ref{fig1},
as the model state for the CCM. We must therefore make a
selection of an appropriate value for $\Phi$. A possible choice
would be the classical ground state of the Hamiltonian of (\ref{ham})
(i.e., $\Phi=\Phi_{\rm cl}$ as given by Eq.~(\ref{phi})).

Another possibility is to perform a CCM-LSUB$n$ approximation calculation
and then to minimize the corresponding LSUB$n$ approximation to the
energy with respect to $\Phi$,
\be \label{phi_lsubn}
 E_{{\rm LSUB}n}(\Phi)\rightarrow\min \quad  \Leftrightarrow \quad \Phi=\Phi_{{\rm LSUB}n} .\ee
The results for $\Phi_{{\rm LSUB}n}$ will be given later (Fig.~\ref{fig_phi}).
However, we note now that although the CCM does not yield a strict upper
bound for the ground-state energy, using
$\Phi=\Phi_{{\rm LSUB}n}$ (i.e., using the CCM with a 
variational parameter) has been found to be 
a reasonable assumption.\cite{Bursill}
There are several additional arguments to suggest that $\Phi=\Phi_{{\rm LSUB}n}$ 
is indeed a better choice than $\Phi=\Phi_{\rm cl}$, as indicated below.

In the first place we note that we cannot find solutions for the LSUB6 equations
using $\Phi=\Phi_{\rm cl}(J')$ in the region $-0.7\lesssim J'\lesssim -0.47$
insofar as the Newton method used to solve these equations does not converge in
that region. This is a clear indication that this model state
is not a good one. By contrast, such behaviour is not found 
for $\Phi=\Phi_{{\rm LSUB}n}$.

Secondly, it is generally known that quantum fluctuations tend to
prefer collinear order\cite{ri93,schultz92} (e.g., N\'eel order).
We will indeed find (and see Fig.~\ref{fig_phi} below) 
that the N\'eel ordering ($\Phi=0$)
seems to survive for some $J'<-1/3$, in which region it has already broken
down in the classical case.
This is also in agreement with results of exact diagonalizations
for our model (and see Fig.~\ref{fig_corr} below).

Thirdly we find better agreement of the CCM results 
for the energy compared to exact diagonalization results by
using the helical state as the model state with the value 
$\Phi=\Phi_{{\rm LSUB}n}$ rather than with the classical value 
$\Phi=\Phi_{\rm cl}$. We find that CCM results for
the ground-state energy usually agree well with the corresponding ED results
(and with results of other methods),\cite{zeng98} provided that a good 
CCM model state is chosen.

We therefore use the helical state with
$\Phi=\Phi_{{\rm LSUB}n}$ as the CCM model state throughout this paper.
Note that for $J'\geq -1/3$ this model state is identical to the classical
ground state of Eq.~(\ref{ham}) but that for $J'<-1/3$ it is not.

\section{Results}
Using the CCM scheme described above, we calculate the 
approximate ground state and the low-lying excitations of the Hamiltonian 
of Eq.~(\ref{ham}). For comparison
we also exactly diagonalize finite sized lattices of square shape.
We use periodic boundary conditions with
$N=16,18,20,26$ and $32$ spins, and we extrapolate to the infinite system
using standard finite-size scaling laws.\cite{neuberger89,schultz92}
We present results for the ground-state energy, the order parameter and
the excitation gap. We examine the formation of local singlets (for $J'>1$),
the effects of frustration (for $J'<0$), and the special case of the
honeycomb lattice ($J'=0$).

\begin{figure}[h]
  \psbild{\centerline{\epsfysize=7cm \epsfbox{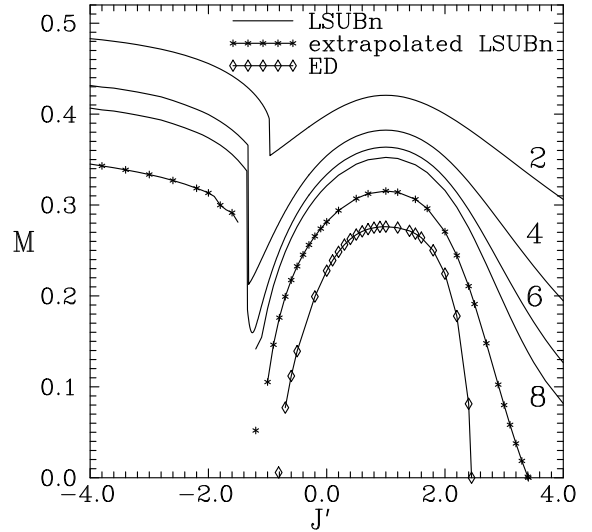}}}
  \vspace{0.3cm}
  \caption{Ground-state magnetic order parameter
           (Eq.~(\protect{\ref{m}})) 
           versus $J'$, for the CCM-LSUB$n$ approximation.
           The results are compared (for the N\'eel region only)
           with $M_s(\infty)$, using
           exact diagonalization (ED) data of the
           antiferromagnetic structure factor, using the ansatz
           $M_s^2=(1/N^2)\sum_{i,j}(-1)^{i+j}\la\Sp_i\cdot\Sp_j\ra=M_s(\infty)^2+\mbox{const}\times N^{-1/2}$.
           Note that both extrapolated results fit poorly in a region
           around $J'\approx-1$, and we therefore plot them here as 
           isolated points (omitting the solid lines).}
  \label{fig_m}
\end{figure}

$J'>1$: {\em Formation of local singlets.} 
Using the CCM we obtain clear indications of a second-order
phase transition to a disordered
dimer-like phase at a certain critical value of 
$J'$, namely $J'_s$. For $J'>J'_s$, 
the N\'eel-like long range order melts, (i.e., the sublattice magnetization,
$M$, given by Eq.~(\ref{m}) becomes zero). Our estimate for $J'_s$
using the four extrapolated LSUB$n$
results for $M$ with $n=2,4,6,8$ (see Fig.~\ref{fig_m}) is $J'_s\approx 3.41$.
However, using only the three CCM LSUB$n$ approximations with $n=4,6,8$
for the extrapolation, we obtain a value $J'_s\approx 3.16$, 
which indicates that the true value could
be even somewhat smaller. This is in agreement with our corresponding 
result using exact diagonalizations of small systems.
By using the extrapolation scheme of Ref.~\onlinecite{schultz92},
we find a critical value $J'_s\approx 2.45$ for the magnetization.
Note however that better accuracy requires larger systems
because of the exact diagonalization (ED) extrapolation 
ansatz for $M$ (i.e., $M=M_{\infty}+{\rm const}\times N^{-1/2}$).
Therefore, we cannot consider the ED results for
the magnetization (and see Fig.~\ref{fig_m}) as quantitatively correct.
Our two results for the critical $J'_s$ also
agree with the estimate $J'_s\approx 2.56$ from series expansion,\cite{singh88}
and even the result $J'_s=3$ from the simple
variational ansatz of Eq.~(\ref{var}) agrees surprisingly well with these values.
By contrast, the second-order renormalized spin wave theory (RSWT)\cite{ivanov96}
gives the larger result $J'_s\approx 5.0$, indicating that the 
standard spin-wave approach is insufficient to describe this type of transition.

\begin{figure}[h]
  \psbild{\centerline{\epsfysize=7cm \epsfbox{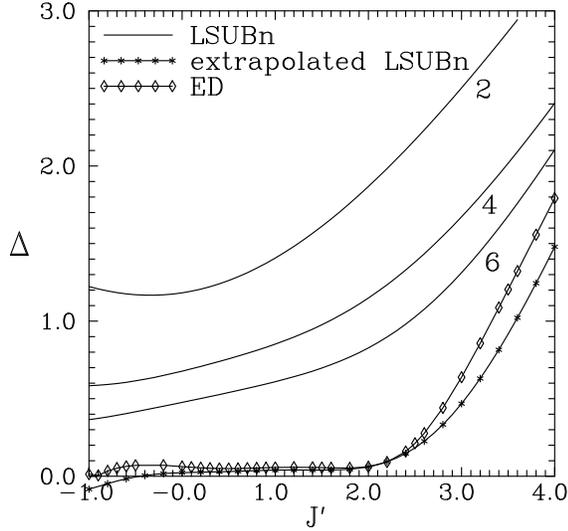}}}
  \vspace{0.3cm}
  \caption{The gap $\Delta$ between the lowest-lying excitation energy and
           the ground-state energy versus $J'$
           using the CCM-LSUB$n$ approximations,
           in comparison with the extrapolated
           result of exact diagonalization (ED) (using the ansatz
           $\Delta=\Delta_{\infty}+\mbox{const}\times N^{-1}$).}
  \label{fig_gap}
\end{figure}

Another indication
of a dimerized phase is the appearance of a gap $\Delta$
between the ground state and the lowest-lying excited state.
We clearly expect a spectrum with gapless Goldstone modes if the ground
state is N\'eel long-range ordered, whereas for a disordered 
singlet ground-state the formation of triplet excitations may cost
a finite amount of energy.
This behaviour is reflected by our results using both CCM and
exact diagonalization (see Fig.~\ref{fig_gap}), which
agree well with each other.
For $J'\gg J'_s$, there is a gap proportional to $J'$, corresponding to the
dimer-like nature of the ground state. 
The gap obviously opens in the range $2.5\lesssim J'_s\lesssim 3.0$ 
in both the ED and CCM calculations. This is in good agreement with the
corresponding estimates for the critical point using the order parameter.
Note, that the standard linear spin wave theory fails in calculating the
gap (i.e., gives gapless modes for all values of $J'$ with $J'>0$).\cite{ivanov96}

\begin{figure}[h]
  \psbild{\centerline{\epsfysize=7cm \epsfbox{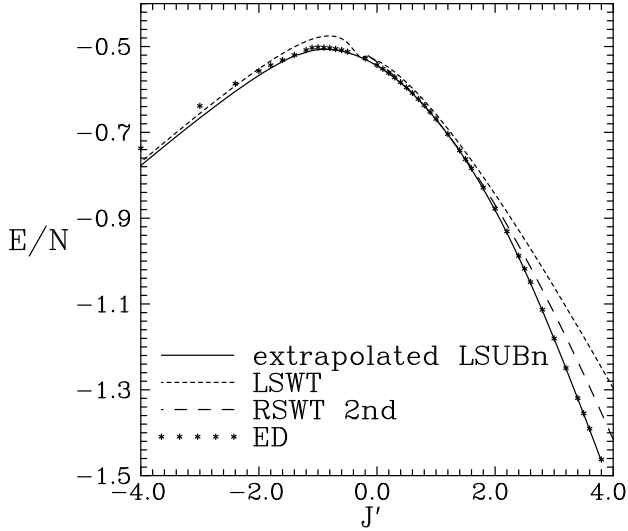}}}
  \vspace{0.3cm}
  \caption{Ground-state energy (Eq.~(\protect{\ref{en}}))
           versus $J'$ for the extrapolated CCM-LSUB$n$ approximations,
           in comparison with results of spin-wave theory
           (linear and second-order renormalized),\protect{\cite{ivanov96}}
           and with the extrapolated result of exact diagonalization (ED) 
           data (using the ansatz
           $E/N=E_{\infty}/N+\mbox{const}\times N^{-3/2}$).}
  \label{fig_energy}
\end{figure}

By comparing the results for the ground-state energy, we find
excellent agreement between the CCM results and the results from exact
diagonalization (see Fig.~\ref{fig_energy}) for $J'>0$.
By contrast, spin-wave theory (SWT) calculations\cite{ivanov96}
show a significant deviation from these results for larger $J'\gg 1$.
These spin-wave results are obviously poor 
since the simple upper bound for the energy given by Eq.~(\ref{en_var})
(e.g., $E_0=-1.5$ for $J'=4$) is smaller than the corresponding 
SWT results (e.g., $E_0=-1.42$ from 2nd-order RSWT). By contrast,
CCM and ED results (both are about $E_0=-1.54$ for $J'=4$) are slightly
smaller than the variational result. While both CCM and SWT
calculations have the N\'eel state as starting point, we find the CCM is much
better able than SWT to describe the transition to the rotationally invariant 
disordered state and to the completely dimerized state (represented
by the variational function of Eq.~(\ref{var}) with $t=1$). 
Note that even the simplest CCM approximation (LSUB2)
gives the correct asymptotic result for the energy
(i.e., Eq.~(\ref{en_var})) for very large values of $J'$, whereas SWT does not.

For the case of the pure square-lattice Heisenberg antiferromagnet
(i.e., $J'=1$), we reproduce the CCM results of
Refs.~\onlinecite{zeng98,bishop99a},
which have already been demonstrated to agree well with those from
other methods.

$J'=0$: {\em honeycomb lattice.}
For the special case of $J'=0$ (which is equivalent to the
honeycomb lattice), we find that the CCM and the ED
results are in good agreement (see Table~\ref{tab2}).
However, the magnetization $M$ for
ED is found to be smaller then the CCM result. We note, however, that the
CCM result for $M$ at this point agrees with the result of high-order
SWT\cite{ivanov96} ($M=0.28$) 
as well as with the result of series expansion\cite{oitmaa92}
($M=0.26$), although it does not agree so well with the result of Monte
Carlo calculations\cite{reger89}
($M=0.22$). We note too that our CCM results here agree perfectly with
previous lower-order CCM calculations.\cite{rosenfeld98}

$J'<0$: {\em Frustration.} 
For $J'\lesssim -2$, we find that the extrapolated ED results for the energy
lie appreciably above the CCM (and SWT) results (see Fig.~\ref{fig_energy}).
This is because the energies
for the small lattices considered do not fit well to the
finite-size scaling law ($E/N=E_{\infty}/N+{\rm const}\times N^{-3/2}$) 
in this region. The finite-size effects for systems 
with an incommensurate helical structure are found to be larger
than for systems with, for example, N\'eel order or with dimerized spin pairs.
However, we find that our best ED result (with 32 spins)
shows only very small deviations from the CCM result, even in the frustrated
region.

While classically we have a second-order phase transition
(from N\'eel order to helical order) at $J'_c=-1/3$, using the CCM we find 
indications for a shift of this critical point to a value $J'_c\approx -1.35$
(see Fig.~\ref{fig_phi}) in the quantum case.
The ED data of the structure factors (see Fig.~\ref{fig_corr}) also show 
a shift of the transition to stronger ferromagnetic $J'$ bonds.
Both correspond to the general picture that quantum fluctuations prefer
a collinear ordering (such as N\'eel order). Hence this ordered state 
can survive for the quantum case
into a region where classically it is already unstable.
The N\'eel model state ($\Phi=0$) gives the minimum ground-state
energy for all values $J'>J'_c$, where $J'_c$ is also dependent on the
level of LSUB$n$ approximation level.
For $J'<J'_c$ another minimum
in the energy for $\Phi \ne 0$ is found to lie lower than the minimum
at $\Phi=0$ (see Fig.~\ref{fig_e}).
The state for $\Phi \ne 0$ is believed to be a quantum analogue of
the classical ground state in two dimensions. Furthermore, the
crossover from one minimum solution to the other is not smooth but
is abrupt at this point (see Fig.~\ref{fig_e} and Fig.~\ref{fig_phi}).
This behaviour is assumed here to be an
indication of a phase transition. Furthermore, it might also
indicate that this is a first-order phase transition
and, consequently, that due to quantum fluctuations
the nature of this phase transition is changed
from the classical second-order type to a first-order type.

The behaviour of the order parameter (see Fig.~\ref{fig_m}) in the region
around $J_c'\approx -1.35$ where we expect the 
above-mentioned quantum phase transition is also quite marked.
We cannot extrapolate the LSUB$n$ results directly, because the
phase transition points shift with the order of the LSUB$n$ 
approximations (Fig.~\ref{fig_phi}).
We thus find a large statistical deviation of 
the extrapolated results in the region $-1.4\lesssim J'\lesssim -1.0$.
Hence, we use the minima for $M$ in that region
to extrapolate an estimation of the order parameter. We find that minimum
($M\approx 0.05$) to be at a value of $J'\approx -1.2$.
The extrapolated ED results do not agree very well with this result,
since these give $M$ to be zero at $J'\approx -0.8$.
However, these results are also very poor in that region
because of the strong influence of the boundary conditions and large
statistical errors.
As a result of these difficulties we are not able to decide whether or not
quantum fluctuations and frustration are able to form
a disordered quantum spin liquid phase (i.e., with $M=0$) 
between the N\'eel state
and the helical state for some finite frustrating $J'<0$ regime.
However, the CCM results suggest that there is either no quantum
spin liquid phase or that, if it does exist, it does so only 
in a very small region.

\section{Summary}
Using the CCM we have studied the influence of quantum spin fluctuations 
on both the ground-state phase diagram and the excited states
of a spin-half square-lattice Heisenberg antiferromagnet
with two kinds of nearest-neighbour exchange bonds. The phase diagram is
found to contain a quantum helical phase, a N\'eel ordered phase, 
and a finite-gap quantum disordered phase. While we have clearly a 
second-order transition from the N\'eel phase to the
finite-gap quantum disordered phase, we also
found indications of a quantum-induced first-order transition from the
N\'eel phase to the helical phase, for which classically we have 
a second-order transition. 
While our CCM results were in general in good agreement with
the ED data, we found the CCM particularly good at describing the dimerized phase.
By contrast, spin wave theory\cite{ivanov96} fails in that region due
to enhanced longitudinal spin fluctuations.
Accurate high-order CCM results for the antiferromagnet on the 
honeycomb lattice were also presented.

\acknowledgments
We would like to thank N.B.~Ivanov for his stimulating discussions.
This work has also been supported in part by the Deutsche
Forschungsgemeinschaft (GRK 14, Graduiertenkolleg on 
``Classification of Phase Transitions in Crystalline Materials''),
and by a research grant (GR/M45429) from the Engineering and
Physical Sciences Research Council (EPSRC) of Great Britain.



\begin{table}
\caption{Number of fundamental ground-state configurations of the LSUB$n$
approximation for the Hamiltonian of Eq.~(\protect{\ref{ham}}), 
using a N\'eel state ($\Phi=0$) and a helical state ($\Phi\neq 0$) for the 
CCM model state, and the number of fundamental excited state 
configurations using the N\'eel model state only.}
\label{tab1}
\begin{tabular}{r|rr|r}
LSUB$n$ & ground state: $\Phi=0$ & $\Phi\neq 0$ & excited state: $\Phi=0$  \\
\tableline
2 &     3 &     5 &    1 \\
4 &    22 &    76 &   16 \\
6 &   267 &  1638 &  331 \\
8 &  4986 & 42160 & 7863 \\
\end{tabular}
\end{table}

\begin{table}
\caption{Ground-state energy per spin, sublattice magnetization, and 
         excitation energy gap for the Hamiltonian of Eq.~(\protect{\ref{ham}})
         with $J'=0$. This special case
         is equivalent to the honeycomb lattice.
         We present (extrapolated) LSUB$n$ results and extrapolated ED 
         results.}
\label{tab2}
\begin{tabular}{r|rrrrr|r}
      & LSUB2 & LSUB4 & LSUB6 & LSUB8 & extrapolated & ED \\
\tableline
$E/N$ & -0.525 & -0.540 & -0.542 & -0.543 & -0.5447 & -0.543 \\
$M$   &  0.399 &  0.354 &  0.334 &  0.321 &  0.28   &  0.23  \\
gap   &  1.182 &  0.678 &  0.476 & --     &  0.02   &  0.06  \\
\end{tabular}
\end{table}

\begin{figure}[h]
  \psbild{\centerline{\epsfysize=7cm \epsfbox{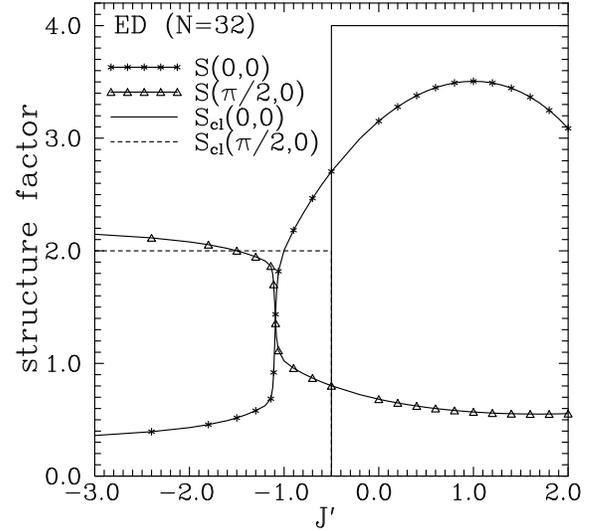}}}
  \vspace{0.3cm}
  \caption{Ground-state structure factor
           $S(\kk)\propto\sum_{i,j\in A}e^{\im(\Rg_j-\Rg_i)\cdot\kk}\la\Sp_i\cdot\Sp_j\ra$
           (i.e., the summation is taken over one sublattice)
           for the Hamiltonian of Eq.~(\protect{\ref{ham}}) with 32 spins,
           for the quantum and the classical case. The N\'eel
           order ($\kk=(0,0)$) becomes unstable against the helical order 
           in the classical model for $J'<-0.5$,
           but in the quantum model the N\'eel
           ordering gives way to helical order only for 
           $J'\protect\lesssim -1.1$
           (i.e., the N\'eel ordering is stable quantum-mechanically
           in a region where it is classically already unstable).}
  \label{fig_corr}
\end{figure}

\begin{figure}[h]
  \psbild{\centerline{\epsfysize=7cm \epsfbox{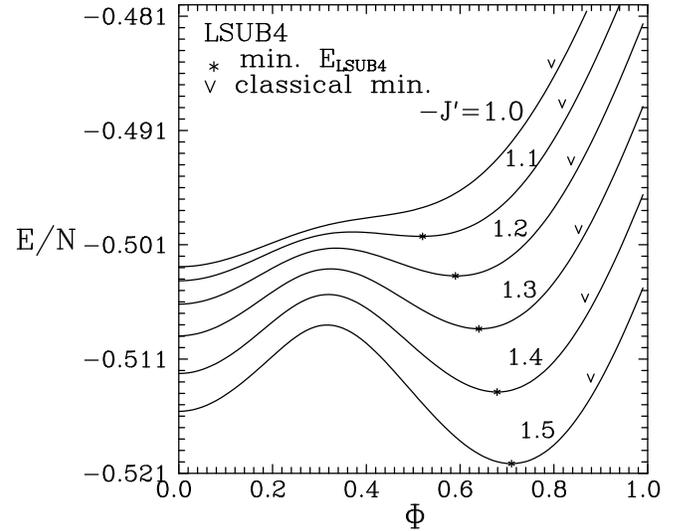}}}
  \vspace{0.3cm}
  \caption{Ground-state energy of the Hamiltonian of 
           Eq.~(\protect{\ref{ham}}) using CCM-LSUB4
           versus the parameter $\Phi$ of the helical CCM model state 
           for certain values of $J'$ in the range 
           $-1.5\leq J'\leq -1.0$. A local minimum
           of $E(\Phi)$ at $\Phi\neq 0$ appears for $J'\protect\lesssim -1.1$,
           which for values of $J'\protect\lesssim -1.35$
           becomes a global minimum (i.e., at $\Phi=\Phi_{{\rm LSUB}4}(J')$),
           indicating the typical scenario of a first-order phase
           transition. The arrows ({\sf v}) indicate 
           the value of $\Phi_{\rm cl}$ for the different values of $J'$.}
  \label{fig_e}
\end{figure}

\begin{figure}[h]
  \psbild{\centerline{\epsfysize=7cm \epsfbox{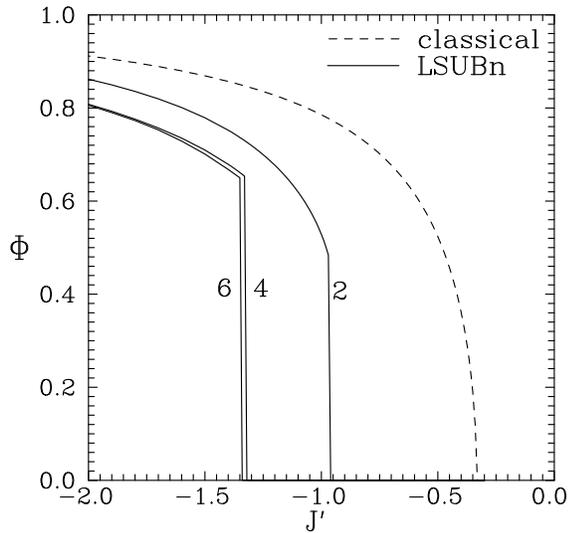}}}
  \vspace{0.3cm}
  \caption{The angle $\Phi_{{\rm LSUB}n}$ (which minimizes the energy $E(\Phi)$,
           see Eq.~(\protect{\ref{phi_lsubn}}))
           versus $J'$, compared with the corresponding classical
           result $\Phi_{\rm cl}$ (see Eq.~(\protect{\ref{phi}})). We find 
           in the quantum case (LSUB$n$) a first-order phase transition 
           (e.g., for LSUB6 at $J'\approx -1.35$ where $\Phi_{{\rm LSUB}n}$ 
           jumps discontinuously from zero to about 0.65).
           By contrast, in the classical case 
           a second-order transition occurs at $J'=-1/3$.}
  \label{fig_phi}
\end{figure}

\end{document}